\documentclass{article}

\usepackage{algorithm}
\usepackage{algpseudocode}
\usepackage{tikz}
\usepackage{arxiv}

\usepackage{amsmath}

\usepackage[utf8]{inputenc} % allow utf-8 input
\usepackage[T1]{fontenc}    % use 8-bit T1 fonts
\usepackage{hyperref}       % hyperlinks
\usepackage{url}            % simple URL typesetting
\usepackage{booktabs}       % professional-quality tables
\usepackage{amsfonts}       % blackboard math symbols
\usepackage{nicefrac}       % compact symbols for 1/2, etc.
\usepackage{microtype}      % microtypography
\usepackage{lipsum}
\usepackage{graphicx}
\graphicspath{ {./images/} }

\title{Cost Transparency of Enterprise AI Adoption}

\author{
  Soogand Alavi \thanks{Equal Contribution}  \\
  Assistant Professor of Marketing\\
  Tippie College of Business\\
  University of Iowa\\
  \texttt{soogand-alavi@uiowa.edu} \\
   \And
 Salar Nozari \footnotemark[1]\\
  Assistant Professor of Marketing\\
  Tippie College of Business\\
  University of Iowa\\
  \texttt{salar-nozari@uiowa.edu} \\
  \And
 Andrea Luangrath \\
 Associate Professor of Marketing\\
  Tippie College of Business\\
  University of Iowa\\
  \texttt{andrea-luangrath@uiowa.edu} \\
}

\begin{document}
\maketitle
\begin{abstract}
Recent advances in large language models (LLMs) have dramatically improved performance on a wide range of tasks, driving rapid enterprise adoption. Yet, the cost of adopting these AI services is understudied. Unlike traditional software licensing in which costs are predictable before usage, commercial LLM services charge per token of input text in addition to generated output tokens. Crucially, while firms can control the \textit{input}, they have limited control over \textit{output} tokens, which are effectively set by generation dynamics outside of business control. This research shows that subtle shifts in linguistic style can systematically alter the number of output tokens without impacting response quality. Using an experiment with OpenAI's API, this study reveals that non-polite prompts significantly increase output tokens leading to higher enterprise costs and additional revenue for OpenAI. Politeness is merely one instance of a broader phenomenon in which linguistic structure can drive unpredictable cost variation. For enterprises integrating LLM into applications, this unpredictability complicates budgeting and undermines transparency in business-to-business contexts. By demonstrating how end-user behavior links to enterprise costs through output token counts, this work highlights the opacity of current pricing models and calls for new approaches to ensure predictable and transparent adoption of LLM services. 
\end{abstract}

% keywords can be removed

\section{Introduction}

Large Language Models (LLMs) are being rapidly adopted in enterprise settings, enabling chatbots, automated content generation, semantic search, coding assistants, and more. As of mid-2025, ChatGPT Enterprise had surpassed 600,000 paying business users, and more than 92\% of Fortune 500 companies were projected to be using OpenAI’s products or APIs in some capacity.\footnote{\url{https://sqmagazine.co.uk/openai-statistics/}} Typically, the cost of adopting traditional software licensing models is predictable and tied to subscriptions or fixed upfront fees. However, the cost of LLM APIs is not fully predictable since they charge per token of text processed as input and tokens generated as output.\footnote{Examples of API pricing based on input and output token counts can be found at \url{https://openai.com/api/pricing/} and \url{https://www.claude.com/pricing\#api}.} While \textit{input} length is in the control of the firm, there is limited control over the \emph{output} tokens, as the model ultimately determines response length and content. Commercial LLM APIs bill input and output tokens, often at different rates, and the true cost of a call is only known once the model completes its response. Consequently, one of the primary cost drivers is opaque and model-determined rather than business-controlled, leaving organizations with expenses shaped by model behavior and user interaction styles. The value derived from digital innovations is shaped not only by their technical capabilities but also by governance mechanisms, pricing models, and process integration \cite{Mojir2022,Lu2021}, and the current lack of control over LLM costs reduces transparency, complicates budgeting, and can diminish the predictability of profit margins at scale for LLM service adopters. 

In this paper, we show that this cost unpredictability can be influenced by latent features of user interaction expressed through linguistic style. Specifically, we reveal how politeness, a linguistic feature of prompt phrasing, influences the number of generated output tokens without altering output quality. Prompts are a form of speech act known as directives \cite{searle1975speech}, and politeness softens these directives. Conventional wisdom suggests that users should prompt LLMs with commands; being direct is a leading recommendation of prompt engineering. Some even advise, ``No need to be polite with LLMs...no need to add phrases like `please', `if you don’t mind', `thank you', `I would like to', etc., and get straight to the point" \cite{bsharat2023principled}. Yet, we reveal that politeness reduces output tokens, thus decreasing enterprise costs.

To identify the causal impact of prompt politeness on generated output length, we develop a methodology to construct counterfactual prompts from a large dataset of real conversations with ChatGPT that vary only in politeness while preserving semantics. Using an experiment that measures the output difference of original and counterfactual prompts, we find that non-polite prompts lead to higher token generation compared to polite prompts. We estimate that this effect alone can be attributed to revenue worth up to \$11M per month for OpenAI.\footnote{Price of output tokens is \$12 per 1M output tokens for GPT4. We find that non-polite prompts lead to more than 14 extra tokens which is equivalent to \$0.000168 extra cost per prompt on average. The average daily queries to OpenAI’s API exceed 2.2 billion. Compared to a scenario in which all prompting is polite, when instead the prompts are non-polite, this generates an additional \$369K revenue per day, simply due to the increase in tokens that non-polite prompts generate in the outcome. This is equivalent to a monthly revenue of \$11M for OpenAI (which is roughly 3\% of its total revenue)} Our findings add nuance to prior industry claims that polite prompts contribute to increased energy usage and LLM providers' monetary costs as those claims neglect the impact on output tokens.\footnote{Sam Altman's response to a question about the use of ``please'' and ``thank you'' in relation to electricity costs can be found at \url{https://x.com/sama/status/1912646035979239430} and \url{https://x.com/tomieinlove/status/1912287012058722659}.} Recent media discussions have focused on how politeness affects the number of input tokens, a cost factor fully under the control of the end user. \footnote{\url{https://medium.com/@ul.tu}} Indeed,  a prompt with ``please do this" requires more input tokens than a prompt with ``do this." Nonetheless, we show that politeness in the input systematically influences the number of output tokens generated, an aspect that depends entirely on the model’s behavior.

Our paper makes three contributions. (1) \textit{B2B implications:} We show that token-based pricing, particularly for generated outputs, can create unmanageable cost expectations for enterprises. (2) \textit{Evidence of uncontrollable cost drivers:} We provide causal evidence that model-determined output length, a first-order factor in API spend, is influenced by end-user style, and is distinct from input length or task complexity. (3) \textit{A measurement framework:} We introduce a counterfactual generation and evaluation pipeline that can isolate causal effects for treatments inside a text by using an LLM to generate the unseen treatment status. 

\section{Background}
\label{sec:background}

\subsection{LLMs in Marketing}

Businesses across industries are increasingly integrating LLM APIs into user-facing services such as LLM-driven chatbots like Intercom's Fin chatbot.\footnote{\url{https://www.intercom.com}} In the marketing literature, LLM adoption is being explored in multiple directions. For example, \cite{goli2024frontiers} evaluate whether LLMs can mimic human survey respondents in intertemporal choice tasks. \cite{ye2025lola} develop and test LLM-assisted online learning algorithms to optimize content delivery, and \cite{li2024frontiers} investigate how large language models can serve as substitutes for human participants in market research. 
Another stream of research has primarily focused on optimizing the output content and accuracy of LLM outputs. \cite{brucks2025prompt} show that the phrasing or structure of a prompt significantly affects how LLMs generate their responses. However, the literature has largely overlooked how prompt phrasing can influence the monetary cost of these outcomes. This gap in the literature is critical since cost variability creates challenges for cost predictability and transparency in B2B contexts, where enterprises require reliable budgeting and billing structures. This study fills in this gap by showing that stylistic prompt features can influence output length and cost paid by the API adopter. 

\subsection{Politeness Theory}

From a linguistic perspective, \textit{how }a prompt is written could potentially impact LLM response. Prompts are a form of speech act known as directives \cite{searle1975speech}. Directives imply that an action must be undertaken by the receiver. In marketing, phrases like ``buy now'' direct and motivate action. Consumers and marketers alike use directives to advise what others should do, buy, or say, and to motivate desired actions. For example, directives have been studied in brand-generated social media posts \cite{villarroel2019cutting}, environmental slogans \cite{kronrod2012eco}, and advertisements \cite{zemack2017just} to motivate consumer behaviors.  

A common strategy to soften a directive in a socially acceptable way is through politeness. Phrases such as ``please buy now'' or ``would you please consider buying?'' offer a more polite request. According to politeness theory \cite{brown1987politeness}, politeness originates out of a concern over the other person and the desire to be liked. Upon initial human encounters, politeness serves as a regulatory tool for smooth social interaction \cite{dufner2023liked} and, in marketing contexts, assists in maintaining satisfactory customer-service provider relationships \cite{coulter2003effects}. Politeness theory resides in a broader domain of relational work \cite{locher2005politeness} on the creation and maintenance of interpersonal relationships. Within this body of work, politeness is considered to be a conventionalized interpersonal script embedded in interaction. Given its fundamental importance in human interactions, it is not uncommon for humans to apply politeness scripts in interactions with LLMs. Yet, our understanding of how politeness in prompts affects LLM response is not well understood.  

\section{Methodology}\label{Method}
We formalize the problem using the framework of potential outcomes. For a prompt denoted by $p_i$ which has a vector of characteristics denoted by $X_i$ and outcome of $Y(p_i)$ defined as the token length of the response from a language model to $p_i$, we first determine the treatment status (T) of $p_i$ by classifying it into two main classes of polite (T=1) and non-polite (T=0). We then construct the unseen status of the prompt through an intervention on the text leading to the generation of the counterfactual prompt, $p_i^{\dagger}$,  with the treatment status (1-T). Finally, to capture the causal estimate of being polite in a prompt, we launch a two-armed digital experiment where we feed the polite prompt (T=1) and the non-polite prompt (T=0) simultaneously into two separate APIs (GPT-4-Turbo)\footnote{We set the temperature parameter to zero and the top p parameter to default for all tasks. Since this setting is set in both control and treatment, we do not expect these parameters to impact the results systematically. } to measure $Y(p_i)$ and $Y(p_i^{\dagger})$. We document details of treatment classification, counterfactual generation, and digital experiment in this section.

\subsection{Data and Treatment Classification}
We use the WildChat dataset \cite{zhao2024wildchat}, a large‑scale corpus of roughly one million real-world conversations between online users and ChatGPT. Each record preserves the original user prompt and the model’s response. Collected by granting free, in-the-wild access to the service, WildChat captures a wide spectrum of naturally occurring interactions that are largely absent from curated instruction-tuning sets. We randomly chose 20K English prompts from the GPT-4 interactions for our study.\footnote{We only used the prompt and discard the output reported in the dataset as we refeed the prompts.} We focused on the first interaction of the user and the model even if the conversation was an ongoing back-and-forth interaction. We first classify these prompts into polite or non-polite prompts. We feed chosen prompts from the WildChat dataset to the GPT-4-Turbo model as a classification task and ask the model to classify them as polite or non-polite. This classification is the assignment of the treatment status for the original prompt corpus. While there are various methods for denoting politeness in text, our main analysis relies on the designation made by the language model itself as the model is the subject of our experiment, and therefore the classification should make sense from the perspective of the model.\footnote{We provide construct consistency with other methods in Section \ref{polite_const}} The \textit{polite} class can either have explicit politeness cues (e.g. inclusion of ``please") or be implicit. The \textit{non-polite} class is then considered anything that is not classified as a polite prompt, even if they are not overtly rude. For our main analysis, we do not require polite prompts to be explicit in form (e.g. explicitly contain the word ``please.'').\footnote{We provide robustness checks based on different forms of politeness in Section \ref{polite_const}}

\subsection{Counterfactual Generation}
Since we are interested in the LLM's behavior in response to politeness in a prompt, we cannot simply model the outcome as a function of treatment and covariates and estimate the effect as in the causal ML literature (see \cite{chernozhukov2016double, kennedy2023towards}). The main obstacle in doing so is that given the context of treatment inside a text, the vector of characteristics, $X_i$, is ultimately a reduction of the information available in the prompt $p_i$ and the size and content of it vary across observed data.  Instead, to enhance the estimation, we aim to recreate the prompt in the unseen treatment status and measure the difference of feeding a pair of prompts that are only different in politeness.

To create the unseen version of the prompts, we need to recreate the same prompt with the opposite tone. We define an intervention where we rephrase the observed text in a (non-polite) prompt into a new (polite) prompt while aiming to keep all other linguistic properties (included in vector $X_i$) the same. Let $T\in[0,1]$ represent the treatment status of a text denoted by $p_i$, $X_i$ a vector of prompt characteristics, and $Y(p_i)$ the downstream outcome of output token length of the LLM response. As treatment is inside a text, we rely on a transformative function to create the counterfactuals. Specifically, we define a textual intervention $f(\,\cdot\,)$ as a point‑wise transformation using a language model to rewrite a text so its overall tone is changed to the other state of treatment (1-T) without otherwise altering other dimensions including its semantic content. For a set of observed prompts $\mathcal{P}$ and a set of generated prompts denoted by $\mathcal{P^{\dagger}}$, function $f(\cdot)$ can be formally denoted as follows:
 
\[
f_{\theta }:\mathcal{P}\longrightarrow \mathcal{P^{\dagger}},
\qquad 
p_i^{\dagger}\mapsto f_{\theta}(p),
\]
where $p_i^{\dagger}$ is the generated counterfactual of $p_i$, and the parameters $\theta$ are defined by an LLM prompt. In our application, $f_{\theta}$ flips the tone of the utterance (polite $\leftrightarrow$ non-polite) while preserving other dimensions. Specifically, for each observed prompt $p_i$ in the \textsc{WildChat} corpus, we construct a counterfactual prompt, denoted by ${p_i^{\dagger}} = f_{\theta}(p_i)$, by prompting GPT-4-Turbo to (i) change the politeness tone and (ii) preserve the semantic meaning and minimize any other change.

It is worth mentioning that as outlined in \cite{banerjee2025language}, in the context of text, it is challenging to keep all dimensions included in vector $X_i$ constant as altering one dimension (T) can potentially change other dimensions too. Our approach of using LLMs to create counterfactuals will ensure minimal changes in semantic structure of the prompt. To check for this, we calculated cosine similarity between original and counterfactual prompts in our dataset, as 0.932, which indicates that prompt characteristics have been greatly preserved. 

One inherent dimension subject to change is the length of the prompt. To mitigate length based confounding, we impose a hard token difference constraint and control for it in our estimation
\[
  \bigl| \operatorname{tokens}(p_i) - \operatorname{tokens}({p_i^\dagger})\bigr|
  \;\le\; \sigma ,
\]

We set $\sigma \leq 5$ in the main specification to give room for all possible variations of the politeness occurring in a prompt\footnote{Specifically, we limit the politeness counterfactual transformation from the minimal change of simply adding (removing) ``please'' (1 token) to more extensive rewrites such as ``could you please [verb] ?'' (5 tokens).}, which reduced the number of prompts in our dataset to 15,961 (prompt pairs).\footnote{Since we do not expect that adding or removing politeness will change the prompt by more than 5 tokens, we manually reviewed 100 excluded pairs and found that the larger token gaps mainly resulted from grammatical errors in the original prompts, which the counterfactuals corrected alongside the politeness transformations. In other words, the increased token differences stem from both grammatical corrections and politeness transformations. These cases were excluded when constructing the final dataset of 15,961 prompt pairs, all of which underwent only the politeness transformation. Notably, the main analyses were also conducted on the full dataset without this exclusion criterion, and the results remained consistent.} Across 15,961 original prompts, 20.8\% were classified as polite. We present examples of original prompts and their corresponding counterfactual versions in Table \ref{example_orig_counter}.

\begin{table}[H]
\centering
\caption{Examples of Polite and Non-Polite Original Prompts with Corresponding Counterfactuals}
\label{example_orig_counter}
\begin{tabular}{|p{3cm}|p{6cm}|p{6cm}|}
\hline
\textbf{Original Tone} & \textbf{Original Prompt} & \textbf{Counterfactual Prompt} \\ 
\hline
Polite & Can you please find me a stevia-free chocolate protein powder? & Find me a stevia-free chocolate protein powder. \\ 
\hline
Non-Polite & Write a critique of the Hundred cricket format & Could you please write a critique of the Hundred cricket format? \\ 
\hline
\end{tabular}
\end{table}

\subsection{Experiment}
The estimation problem using text transformation is then a straightforward generalization of the potential outcomes framework, where instead of the treatment T, we are interested in a transformation of a treatment, the text, which has the treatment embedded inside. This transformation leads to an effect estimator for a general class of text transformations, like rephrasing with politeness. For any realized prompt $p$, the potential outcome is $Y(p)$; the counterfactual if we rewrote that same prompt with the opposite tone is $g_\theta(p)$ with the outcome of $Y\!\bigl(g_\theta(p)\bigr)$.  
\[
Y = 
\begin{cases}
    Y(p) & \text{if } T \text{ is observed},\\
    Y(f_{\theta}(p)) & \text{if } 1-T \text{ is generated}.
\end{cases}
\]

Having the counterfactuals, we feed the control prompts (T=0) to a GPT-4-Turbo and the treatment prompts (T=1) to the exact same model using a separate API. For each prompt $p_i$ we record $Y ( p_{i})$ defined as the number of output tokens generated by the model in response to prompt $p_i$ (as reported by the API). The difference of tokens for each pair of prompts will therefore be measured as $$\tau_i(p)=\; 
  \mathbb{E}\!\bigl[Y\!\bigl(p_i\bigr)\bigr] -
  \mathbb{E}\!\bigl[Y(f(p_i))\bigr]$$
  which then leads to ATE,
\[
\bar\tau(p)\;=\;\frac{1}{k}\sum_{i=1}^k\tau_i(p)
\] 

The summary statistics in Table \ref{tab:summary_stats_input_output} show that the average number of output tokens in the treatment (polite) is lower than the control (non-polite).

\begin{table}[H]
\centering
\small
\caption{Summary Statistics}
\label{tab:summary_stats_input_output}
\begin{tabular}{p{6cm}rrr}
\hline
\textbf{Category} & \textbf{Count} & \textbf{Mean} & \textbf{Std. Dev.} \\
\hline
\multicolumn{4}{l}{\textbf{Input Token Characteristics}} \\
\textit{Control (Polite = 0)} & 15961 & 31.528 & 26.731 \\
\textit{Treatment (Polite = 1)} & 15961 & 33.529 & 26.170 \\
\hline
\multicolumn{4}{l}{\textbf{Output Token Characteristics}} \\
\textit{Control (Polite = 0)} & 15961 & 495.740 & 292.491 \\
\textit{Treatment (Polite = 1)} & 15961 & 483.352 & 292.810 \\
\hline
\end{tabular}
\end{table}

Ultimately, we estimate the impact of politeness on output token length using the following equation.

$$
\text{OutputTokenLength} = b_0 + b_1 \cdot \text{Treatment} + b_2 \cdot \text{InputTokenLength}
\label{eq:polite_model}
$$

\section{Results}\label{Results}
The main estimation results as shown in Table \ref{tab:output2}, suggests that using a polite prompt reduces the output length by 14.426 tokens. In the following subsections, we provide several robustness checks.

\begin{table}[H]
\centering
\small
\caption{Estimation Results for the Impact of Polite Prompt on Generated Output Length}
\label{tab:output2}
\begin{tabular}{lrr}
\hline
\textbf{Variable} & \textbf{Coef.} & \textbf{Std. Err.} \\
\hline
Intercept & 463.626$^{***}$ & 3.017 \\
Polite & -14.426$^{***}$ & 3.264 \\
Input Tokens & 1.019$^{***}$ & 0.062 \\
\hline
\multicolumn{3}{l}{Adjusted R-squared = 0.0089} \\
\multicolumn{3}{l}{Number of Observations = 31,922} \\
\hline
\multicolumn{3}{l}{\textit{Note:} $^{***} p<0.01$, $^{**} p<0.05$, $^{*} p<0.1$} \\
\hline
\end{tabular}
\end{table}

\subsection{Robustness to Politeness Construct}\label{polite_const}
Politeness in a prompt can be explicit, such as using words like please and thank you, or it can be implicit. In Table \ref{tab:politetype}, we show that the results hold for all subsets of the data in which the politeness is explicit or implicit. In column (1), the subset includes prompts classified as polite due to the presence of ``please" or ``thank you", and column (2) focuses on prompts that only include ``please". The remaining polite classifications that do not fall into either column (1) or (2) are grouped under column (3) and include implicit forms of politeness such as common phrases like ``can you", or ``could you".

\begin{table}[H]
\centering
\small
\caption{Estimation Results Based on Politeness Types}
\label{tab:politetype}
\begin{tabular}{lccc}
\hline
\textbf{Variable} & \textbf{(1)} & \textbf{(2)} & \textbf{(3) } \\
\textbf{} & \textbf{Please and Thank you} &  \textbf{Please} & \textbf{Implicit}\\
\hline
Intercept & 486.441$^{***}$ & 486.956$^{***}$ & 412.910$^{***}$ \\
 & (3.818) & (3.820) & (4.829) \\
Polite & -16.069$^{***}$ & -16.060$^{***}$ & -10.882$^{*}$ \\
 & (4.187) & (4.189) & (5.083) \\
Input Tokens & 0.865$^{***}$ & 0.853$^{***}$ & 1.482$^{***}$ \\
 & (0.082) & (0.082) & (0.091) \\
\hline
Adjusted R-squared & 0.0058 & 0.0056 & 0.0232 \\
Number of Observations & 20,798 & 20,780 & 11,142 \\
\hline
\multicolumn{4}{l}{\textit{Note:} $^{***}p<0.01$, $^{**}p<0.05$, $^{*}p<0.1$} \\
\hline
\end{tabular}
\end{table}

As an additional check, we also construct a politeness measure using LIWC \cite{boyd2022development}. This robustness check was conducted for two main reasons: (1) LIWC is a deterministic, steady-state model such that an evaluation of politeness produces the same result every time on the same text, unlike GPT, which is a probabilistic model, therefore allowing us to measure its alignment with the GPT-based polite classification and (2) the politeness measure produced by LIWC is a continuous measure, allowing us to capture the degree of politeness represented in a prompt.  

To achieve the first goal, we label a prompt as polite if its LIWC polite score is greater than 0, and label it as non-polite if the LIWC polite score is equal to 0. Using this label, we calculate an alignment rate between GPT-based classification and the LIWC-based classification, finding an alignment of 81\%. It is important to note that this value does not represent a classification accuracy score, but rather the degree of consistency between our GPT-based politeness classification and the LIWC-derived measure. Further, we construct a subset of the dataset in which the GPT-based and LIWC-based politeness classifications align and re-estimate our model on this restricted sample. As reported in column (1) of Table~\ref{tab:politeliwcest}, a polite prompt is associated with a reduction in output length of approximately 14 tokens on average.

For the second goal, since the LIWC framework measures politeness as a continuous variable, we additionally estimate the model using this continuous politeness measure. The results, as shown in column (2) of Table~\ref{tab:politeliwcest}, indicate that holding all other factors constant, a one-unit increase in the LIWC politeness score corresponds to an average decrease of approximately 5 output tokens.

\begin{table}[H]
\centering
\small
\caption{Estimation Results on LIWC Classification Alignment Subset}
\label{tab:politeliwcest}
\begin{tabular}{lcc}
\hline
\textbf{Variable} & \textbf{(1) Polite as Binary} & \textbf{(2) Polite as Continuous} \\
\hline
Intercept & 472.270$^{***}$ & 489.149$^{***}$ \\
 & (3.339) & (3.314) \\
Polite & -14.850$^{***}$ & -5.851$^{***}$ \\
 & (3.640) & (0.414) \\
Input Tokens & 0.978$^{***}$ & 0.714$^{***}$ \\
 & (0.068) & (0.070) \\
\hline
Adjusted R-squared & 0.0084 & 0.0154 \\
Number of Observations & 25,906 & 25,906 \\
\hline
\multicolumn{3}{l}{\textit{Note:} $^{***}p<0.01$, $^{**}p<0.05$, $^{*}p<0.1$} \\
\hline
\end{tabular}
\end{table}

\subsection{Robustness to Prompt Task Type}
We further test whether there were heterogeneous effects based on task type. We use a semantic similarity approach to classify prompts into predefined task categories. Specifically, we define a set of representative task classes (e.g., information seeking, text generation, editing and rewriting, classification, summarization, technical tasks), each described by a short textual definition. Using the SentenceTransformer model (all-MiniLM-L6-v2), both the class descriptions and prompts are transformed into semantic embeddings that capture their underlying meanings. Cosine similarity is then computed between each input embedding and the class embeddings, and the task label corresponding to the highest similarity score is assigned to the prompt.  

First, we add the prompt task category as a control vector for our estimation, to show the robustness of the main effect after controlling for the prompt task. The results are shown in column (1) of Table \ref{tab:promptclass_interaction}. Additionally, we include the interaction of the task with the treatment variable, to identify any heterogeneous effects.  The results are shown in column (2) of Table \ref{tab:promptclass_interaction}. As seen in the estimation results, polite prompts reduce output length tokens even after controlling for the prompt task and show no heterogeneous effects.

\begin{table}[H]
\centering
\small
\caption{Estimations with Prompt Task Controls}
\label{tab:promptclass_interaction}
\begin{tabular}{lcc}
\hline
\textbf{Variable} & \textbf{(1)} & \textbf{(2)} \\
\hline
Intercept & 381.674$^{***}$ & 384.682$^{***}$ \\
 & (4.809) & (6.021) \\
Polite & -14.170$^{***}$ & -20.685$^{*}$ \\
 & (3.116) & (8.286) \\
Technical Tasks & 112.063$^{***}$ & 106.155$^{***}$ \\
 & (6.028) & (8.253) \\
Editing \& Rewriting & -46.460$^{***}$ & -56.049$^{***}$ \\
 & (6.601) & (9.319) \\
Information Seeking & 36.622$^{***}$ & 28.122$^{***}$ \\
 & (5.504) & (7.767) \\
Summarization & 4.186 & 3.239 \\
 & (6.467) & (9.036) \\
Text Generation & 190.341$^{***}$ & 190.786$^{***}$ \\
 & (4.892) & (6.804) \\
Polite $\times$ Technical Tasks &  & 12.630 \\
 &  & (12.022) \\
Polite $\times$ Editing \& Rewriting &  & 19.186 \\
 &  & (13.209) \\
Polite $\times$ Information Seeking &  & 16.760 \\
 &  & (11.003) \\
Polite $\times$ Summarization &  & 2.195 \\
 &  & (12.937) \\
Polite $\times$ Text Generation &  & -0.492 \\
 &  & (9.789) \\
Input Tokens & 0.920$^{***}$ & 0.923$^{***}$ \\
 & (0.060) & (0.060) \\
\hline
Adjusted R-squared & 0.0987 & 0.0988 \\
Number of Observations & 31,922 & 31,922 \\
\hline
\multicolumn{3}{l}{\textit{Note:} $^{***}p<0.01$, $^{**}p<0.05$, $^{*}p<0.1$} \\
\hline
\end{tabular}
\end{table}

\subsection{Robustness to the Quality of Output}
While prior analyses demonstrated that polite prompts decrease the number of output tokens, a logical next question is whether this effect is due to the differences in quality of the output generated from polite versus non-polite prompts. If output quality remains constant, then it stands to reason that politeness is a cost-reduction strategy for businesses adopting LLM APIs. To address this question, we proceed with two separate approaches.

\subsubsection{Evidence from Data}
First, we show that the outputs generated from the original and counterfactual prompts are semantically similar. To evaluate this, we measure the semantic similarity between each original and counterfactual model output to assess how closely their meanings align. Using the sentence transformer model of all-MiniLM-L6-v2, we embed each text into a high-dimensional semantic vector space, where sentences with similar meanings occupy nearby positions. For each paired output, we compute the cosine similarity between their embeddings, which could potentially range from 0 (no semantic overlap) to 1 (identical meaning). This metric captures semantic relatedness beyond surface-level word overlap. Following common practice derived from SBERT’s correlation with the Semantic Textual Similarity Benchmark \cite{reimers2019sentence}, a cosine similarity greater than 0.8 typically indicates paraphrase, while scores between 0.7 and 0.8 reflect strong semantic similarity. Hence, our average cosine similarity of 0.78 between the original and counterfactual generated outputs indicates that the pairs are strongly similar and close to paraphrastic in meaning.

Next, we explore whether there are systematically common words that are removed when the output is generated with a different tone. We find that the most frequently occurring removed words are common stop words (e.g. have, more, other, not, where, an, like, at, and into). These words generally function as grammatical connectors or modifiers rather than carrying substantive semantic meaning. As an additional check, we remove the stopwords since we suspect that stop words are simply frequently used words that appear with longer sentences. By doing so, we can explore whether the difference in the generated tokens is driven by repeated phrases or words that carry semantic meaning. To examine this, we analyze phrases ranging from one to four words to identify whether they are removed when the input tone is changed. We do not find any specific or meaningful set of phrases that are systematically being used or removed in generated outcomes between the polite and non-polite conditions. 

\subsubsection{Evidence from Experiment}\label{quality_exp}
As a second approach to evaluating output quality across the treatment and control conditions, we rely on human evaluations. From our full prompt-response dataset, we randomly selected 20 prompt pairs (20 polite and 20 non-polite) after excluding prompts whose content involves technical, computational, political, and sensitive or inappropriate topics. The design of this study was a 2 (polite prompt vs. non-polite prompt) $\times$ 2 (original prompt vs. counterfactual prompt) manipulated between subjects. To ensure that the quality does not differ, we randomly assigned participants to one of these four conditions: polite original prompt, polite counterfactual prompt, non-polite original prompt, or non-polite counterfactual prompt.

Prolific participants ($N = 401$; $M_{age} = 42.72$, 53.9\% female, 45.6\% male, 0.2\% prefer not to say, 0.2\% not reported) completed the study for payment and were randomly assigned to view a prompt and the corresponding output to the prompt. Participants were asked to respond to one question rating the quality of the response with the item, ``The response from ChatGPT is high quality,'' on a 7-point Likert scale (1 = strongly disagree, 7 = strongly agree). A two-way ANOVA on quality perceptions revealed an insignificant main effect of politeness ($F(1, 397) = .651$, $p = .420$, $\eta_p^2 = .002$), an insignificant main effect of the origin of the prompt ($F(1, 397) = 2.203$, $p = .139$, $\eta_p^2 = .006$), and an insignificant interaction between the two ($F(1, 397) = 2.103$, $p = .148$, $\eta_p^2 = .005$). 

Participants did not perceive quality differences in the output generated by ChatGPT based on whether the response was to an original prompt ($M = 5.42$, $SD = 1.31$) or one that was constructed as a counterfactual ($M = 5.21$, $SD = 1.51$). This provides additional assurance that our counterfactual generation process did not augment the prompts in a systematic way that affected the quality of the output. Importantly, participants did not perceive quality differences in the output generated by ChatGPT in response to polite ($M = 5.26$, $SD = 1.49$) or non-polite prompts ($M = 5.37$, $SD = 1.35$). This suggests that the reduction in output tokens is not due to a loss of output quality. 

In conclusion, the additional costs that businesses incur when generating outputs from non-polite prompts appear to be largely driven by paraphrased varied content, rather than substantive differences in meaning.

\section{Discussion and Implications}\label{implications}
Given the current pricing structure for adopting LLM services, the results of this paper demonstrate how subtle linguistic features can systematically affect how much an enterprise pays, ultimately revealing that linguistic shifts can balloon revenues for LLM service providers, like OpenAI. One obvious solution could be to explicitly instruct the model on how many tokens to generate. However, LLMs are not particularly adept at self-regulating output length.\footnote{See OpenAI Community Notes: \url{https://community.openai.com/t/gpt-cannot-count-words-why/996739}} To illustrate this, we tested how closely the LLM adheres to length specifications. We used one of our dataset prompts, ``What is earthing?'', and appended the phrase ``describe with \textit{z} number of tokens.'' We tested five different values of \textit{z} ranging from 100 to 300 in increments of 50 (i.e., 100, 150, 200, 250, and 300 tokens). None of the generated outputs matched the exact number of tokens requested; instead, the outputs contained 118, 152, 211, 237, and 306 tokens, respectively. In most cases, the model produced more tokens than specified. This calls into question whether prompting is a reliable mechanism for cost control in business contexts where expenses are tied directly to output length. Another potential solution is to use the \textit{max\_count} parameter of the API\footnote{\url{https://help.openai.com/en/articles/5072518-controlling-the-length-of-openai-model-responses}} to limit the maximum number of tokens (and thus the maximum cost), this parameter has a key drawback that it may prematurely truncate the output once the limit is reached, leaving the response incomplete. Moreover, it does not offer precise control over the exact number of tokens generated, and therefore the exact cost remains unpredictable, but rather it can provide an upper bound of expenses.

While the solution to this problem is currently challenging, policymakers could push for transparency and controllability over output token generation and its cost. When utilizing LLM-based services, a more sophisticated pricing approach is needed so that businesses can predict their adoption costs. While challenging, a potential solution can be to charge API adopters for the lowest possible number of output tokens for a given prompt, providing assurance that they have not been overcharged. Businesses, LLM service providers, and policymakers should require cost transparency of LLM services; the scope, scale, and urgency of output-token cost transparency cannot be understated as businesses implement AI solutions at scale.

\section{Conclusion}\label{conclude}

This research illuminates an output-token problem in which the cost of AI adoption for an enterprise is unpredictable, uncontrollable, and set by the model’s response behavior, rather than being set by the firm. LLM services charge per input and output token; while firms often control the \textit{input}, they have limited influence over \textit{output} tokens, as these are ultimately generated by the model's dynamics. Crucially, this research reveals that subtle shifts in linguistic style affect output token length. Conventional wisdom suggests that prompt politeness is unnecessary when interacting with LLMs. In contrast, our work demonstrates that non-polite prompts increase output tokens, generating additional costs for enterprise AI adopters.

The contribution of this manuscript is both methodological and substantive. Methodologically, building on common practices of analyzing text to generate insights \cite{berger2020uniting}, this paper introduces a framework that leverages LLMs to facilitate causal inference for text-based treatments. Our approach uses LLMs to generate counterfactual text variations that maintain key contextual and stylistic elements of the original documents, enabling causal estimation of treatments within text. 

Substantively, while most of what is considered in AI research today is the curation of algorithms for enhancing the \textit{quality} of AI-generated output, we take a different approach by revealing the lack of transparency in the cost incurred when generating output, even after quality is held constant. While prior work has addressed pricing transparency with market participants \cite{li2025endogenous} as well as consumers \cite{mohan2020lifting}, we document the issue of cost transparency within AI systems, in which subtle linguistic shifts can dramatically affect output token length, and thus create non-transparent enterprise costs. This work provides cautionary guidance to B2B firms employing enterprise AI systems.

% References
\newpage
\addcontentsline{toc}{chapter}{Bibliography}

\begingroup
\small
\bibliographystyle{unsrt}
\bibliography{reff}
\endgroup

\end{document}